\newcommand\lsim{\mathrel{\rlap{\lower4pt\hbox{\hskip1pt$\sim$}}
    \raise1pt\hbox{$<$}}}
\newcommand\gsim{\mathrel{\rlap{\lower4pt\hbox{\hskip1pt$\sim$}}
    \raise1pt\hbox{$>$}}}

\documentstyle[epsf  
]{mn}

\begin{document}
\title{Cluster abundance and large scale structure}

\author[Jiun-Huei Proty Wu]
{Jiun-Huei Proty Wu\thanks{e-mail: jhpw@astro.berkeley.edu}\\
  Astronomy Department,
  University of California, Berkeley,
  601 Campbell Hall, Berkeley, CA 94720-3411, USA
}

\maketitle

\begin{abstract}
We use the presently observed number density of large X-ray clusters
and the linear mass power spectra from galaxy surveys
to constrain the amplitude of matter density perturbations
on the scale of $8 h^{-1}$Mpc ($\sigma_8$),
and the redshift distortion parameter ($\beta$), in both
open cosmologies and flat models with a non-zero cosmological constant.
The best fit to the observed mass power spectra gives
$n=0.84\pm 0.67$ and $\Gamma=0.27^{+0.42}_{-0.16}$,
with the theoretically expected degeneracy
$\Gamma'=0.247\Gamma\exp(1.4n)=0.220^{+0.036}_{-0.031}$
 (all at 95 per cent confidence level).
These are consistent with the recent CMB results. 
Based on this, 
we then calculate the cluster-abundance-normalized $\sigma_8$,
using different models of mass function.
The models considered are by Press \& Schechter (PS; 1974),
by Sheth \& Tormen (ST; 1999),
and
by Lee \& Shandarin (LS; 1999).
The last two
incorporate non-spherical gravitational collapse,
and the $\sigma_8$ based on these two models are significantly lower. 
This lower normalization results from the larger
mass function within the scale range of our interest.
In particular,
we combine the results of these two models to yield
$\sigma_{8{\rm (ST+LS)}}=0.477\Omega_{\rm m0}^\alpha$,
where
$\alpha=-0.3-0.17\Omega_{\rm m0}^{0.34}-0.13\Omega_{\Lambda 0}$.
In our analysis,
we also derive the probability distribution function
of cluster formation redshift
using the Lacey-Cole formalism (1993),
but with modifications to incorporate non-spherical collapse. 
The origins of uncertainties in our $\sigma_8$ results are also 
investigated separately,
with the main contributer being the normalization in the virial mass-temperature relation.
From the PSCz power spectrum alone and
using $\Gamma'=0.220^{+0.036}_{-0.031}$ as the prior,
we also obtain for the IRAS galaxies $\sigma_{8\rm (I)}=0.78\pm 0.06$ 
(at 95 per cent confidence level) . 
By combining this with the $\sigma_8$ result,
we are able to constrain the redshift distortion parameter 
$\beta_{\rm I}$,
which is in turn lower in the non-spherical-collapse models. 
We found
$\beta_{\rm I(ST+LS)}=0.613\Omega_{\rm m0}^{0.24-0.16(\Omega_{\rm m0}+\Omega_{\Lambda 0})}$.
This is more consistent with the recent observations than the result based on the PS formalism.
\end{abstract}

\begin{keywords}
galaxies: clusters -- large-scale structure of Universe 
  -- X-rays -- cosmology: theory
\end{keywords}


\section{Introduction}
\label{intro}

One of the most important constraints on models of structure formation is the 
observed abundance of galaxy clusters. 
Because they are the largest virialized objects in the universe,
their abundance can be simply predicted by linear perturbation theory.
In the literature, 
the cluster abundance has been widely used to constrain
different cosmological models
\cite{White,ECF,VL2,Kitayama,WS,AWS}.
Here,
we shall consider
its constraint on
the amplitude of matter density perturbations
on the scale of $8 h^{-1}$Mpc, $\sigma_8$,
and the redshift distortion parameter, $\beta$, 
in the standard inflationary models.
This requires the knowledge about
the linear power spectrum of mass perturbations,
the so-called `mass function'
(the differential number density of collapsed objects as a function of mass),
the probability distribution function of cluster formation redshift,
and
the virial mass-temperature relation.
Although this has been studied by many other authors
\cite{VL2,ECF,WS,Borgani,PSW},
we revisit this problem
with more caution
to incorporate the recent progress
in measuring the linear mass power spectrum,
as well as understanding the mass function.
Based on these new developments,
we shall investigate $\sigma_8$ and $\beta$
in both
open cosmologies and flat models with a non-zero cosmological constant.

The structure of the paper is as follows.
In section~\ref{powerspectrum},
we first investigate the observed linear mass power spectra.
This includes the estimation of $\sigma_8$,
the spectral index $n$,
and the shape parameter $\Gamma$.
They parameterize the linear mass power spectrum 
predicted by the standard inflationary models.
Within such models,
we reveal the degeneracy between $n$ and $\Gamma$,
and thus reparameterize the linear mass power spectrum
with only a single parameter,
the degenerated shape parameter $\Gamma'$.
The IRAS $\sigma_{8{\rm (I)}}$ is also estimated here.
In section~\ref{the-mass-function},
we explore different models of mass function.
The models considered are
by Press \& Schechter (PS; 1974),
by Sheth \& Tormen (ST; 1999),
and
by Lee \& Shandarin (LS; 1999),
the last two of which incorporate non-spherical collapse.
In section~\ref{formation-redshift},
we then derive
the probability distribution function of cluster formation redshift
for different models of mass function.
The formalism employed is by Lacey \& Cole (1993, 1994),
but modified to incorporate non-spherical collapse.
In section~\ref{Mass-temperature},
we describe how the virial mass of different formation redshift
is related to its temperature,
and lay down the formalism
that relates $\sigma_8$ to the cluster abundance.
In section~\ref{results-and-discussion},
we first calculate the $\sigma_8$ normalized to the observed cluster abundance.
We present the estimates of the $\sigma_8$
based on different models of mass function,
and compare these results with other work in the literature.
The uncertainties of our results are investigated,
with an explicit form for the dependence on various error sources.
We further show how to derive $\beta$ from the information we have,
and compare our results with recent observations.
Finally, in section~\ref{conclusion},
we give a brief conclusion.

\section{Matter perturbations and power spectrum}
\label{powerspectrum}

The standard deviation of matter density perturbations at a smoothing scale $R$
is related to the mass power spectrum ${\cal P}(k)$ as
\begin{equation}
  \label{sigma_R2}
  \sigma^2(R)
  =
  \int |w(kR)|^2 S(k) \frac{dk}{k},
\end{equation}
where
$w(x)$ is the Fourier transform of an unit top-hat spherical window:
\begin{equation}
  w(x)=\frac{3(\sin x-x\cos x)}{x^3},
\end{equation}
and
\begin{equation}
  \label{Sk}
  S(k)
  =
  \frac{k^3}{2\pi^2} {\cal P}(k),
\end{equation}
which is dimensionless and thus independent of the units of $k$.
We shall use
$h^{-1}$Mpc and $h$Mpc$^{-1}$ as the units of $R$ and $k$ respectively,
where $h$ is the present Hubble parameter $H_0$
in units of $100$ km s$^{-1}$Mpc$^{-1}$.
A subscript `0' will denote a quantity evaluated at the present epoch.
Theoretically,
the matter power spectrum ${\cal P}(k)$ in adiabatic inflationary models 
can be expressed as
\begin{equation}
  \label{Pk}
  {\cal P}(k)\propto k^n T^2(k),
\end{equation}
where $n$ is the `spectral index',
which specifies the scale dependence of the initial perturbations,
and $ T(k)$ is the `transfer function',
which transfers the initial perturbations to the present epoch.
An accurate and analytically motivated form of $T(k)$ is
\cite{EisensteinHu98}
\begin{equation}
  \label{Tk}
  T(k)=
  \left\{
    1+\frac{\left[14.2+731/(1+62.5q)\right]q^2}{\textrm{ln}(2e+1.8q)}
  \right\}^{-1},
\end{equation}
where $q=k/\Gamma$
($q$ is originally defined in units normalized to 
the inverse of the horizon size
at the epoch of radiation-matter-energy-density equality), 
and \cite{Sugiyama95}
\begin{equation}
  \label{Gamma}
  \Gamma=\Omega_{\rm m0} h
  \exp(-\Omega_{\rm B0}-\frac{\Omega_{\rm B0}}{\Omega_{\rm m0}})\,.
\end{equation}
Here $\Omega_{\rm m0}$ and $\Omega_{\rm B0}$ are the present
matter and baryon energy densities respectively.
For a fixed $n$,
$\Gamma$ determines the location of the broad peak in ${\cal P}(k)$,
and thus the name `shape parameter'.
In the analysis of later sections,
we shall use $\Omega_{\rm B0}=0.05$,
and then investigate the dependence of our final results on this.

\begin{figure}
  \centering 
  \leavevmode\epsfxsize=8cm \epsfbox{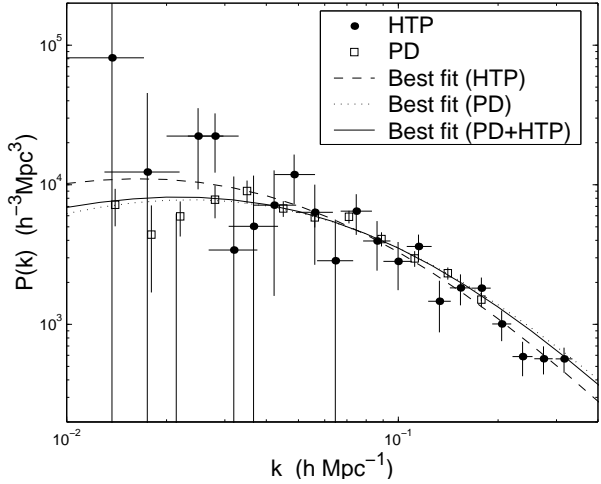}\\
  \caption[]
  {Mass power spectra of different observations and their best fits
   (see text for details).}
  \label{figure1}
\end{figure}

From the theoretical modeling (\ref{Pk}),
we see that the shape of ${\cal P}(k)$ has two essential parameters,
$n$ and $\Gamma$,
i.e., ${\cal P}(k)\equiv {\cal P}(k;n,\Gamma)$.
Thus a comparison between this modeling and observations will give us 
some estimates of $n$ and $\Gamma$.
The observations we considered are 
the linear mass power spectrum ${\cal P}_{\rm (PD)}(k)$ 
by Peacock and Dodds (1994; hereafter PD),
which was compiled from surveys of 
different classes of galaxies and galaxy clusters,
and 
the decorrelated linear IRAS galaxy-galaxy power spectrum ${\cal P}_{\rm I(HTP)}(k)$
by Hamilton, Tegmark, and Padmanabhan (2000; hereafter HTP),
which was based on the IRAS Point Source Catalogue Redshift Survey 
(PSCz; Saunders et al.\ 2000).
As advised by the authors,
to avoid the effect of non-linear evolution on smaller scales (larger $k$),
we used only the first twelve data points of PD ($k\lsim 0.2 h$Mpc$^{-1}$),
and the first twenty two data points of HTP ($k\lsim 0.3 h$Mpc$^{-1}$).
They are shown in figure~\ref{figure1}
as squares and dots respectively (with associated error bars).
Using the maximum-likelihood method based on a $\chi^2$ analysis,
the results of the best estimates
are shown in table \ref{table1} and figure \ref{figure1}.

In table~\ref{table1},
all the errors are at 95 per cent confidence level.
The last column gives the $\chi^2$ values of the best fits,
the degrees of freedom (DOF) in the analysis, 
and their corresponding confidence levels (CL).
In obtaining the results for the HTP (the bracketed numbers),
we have used a top-hat prior $2\geq n \geq 0$
(see later for reasons).
In figure \ref{figure1},
the best fits of ${\cal P}(k)$ for PD and HTP
are shown as the dotted and dashed lines respectively.
When compared with the HTP (dots and dashed line),
the more constraining feature in the PD data (squares) on larger scales (smaller $k$)
tends to confine the peak of the best-estimated ${\cal P}(k)$ (dotted line)
to smaller scales (larger $k$),
and thus a larger estimate of $\Gamma$ (see table~\ref{table1}).

\begin{table}
  \centering 
  \begin{tabular}{c|cccccc}
    \hline\hline
    & $n$ & $\Gamma$ & $\Gamma'$ & $\chi^2$/DOF (CL) \\
    \hline
    HTP    &  $\!\!\! \left(0.91^{+1.09}_{-0.91}\right)$ & $\!\!\! \left(0.18^{+0.74}_{-0.18}\right)$ 
           & $\!\!\! 0.160^{+0.085}_{-0.051}$ & 15.4/19 (70\%) \\
    \hline
    PD     & $\!\!\! 0.99^{+0.81}_{-0.86}$ & $\!\!\! 0.23^{+0.55}_{-0.16}$
           & $\!\!\! 0.229^{+0.042}_{-0.033}$ & 6.95/9 (64\%) \\
    \hline
    $\!\!\!$HTP+PD & $\!\!\! 0.84^{+0.67}_{-0.67}$ & $\!\!\! 0.27^{+0.42}_{-0.16}$
           & $\!\!\! 0.220^{+0.036}_{-0.031}$ & 24.6/30 (74\%) \\
    \hline\hline
  \end{tabular}
  \caption{Best fits of different LSS data sets.
    The errors are at 95 per cent confidence level.
    The bracketed numbers were obtained using a prior $2\geq n \geq 0$.
    See text for more details.}
  \label{table1}
\end{table}

We then combined both the HTP and PD
into the same likelihood analysis (denoted as HTP+PD)
with all data points equally weighted.
In the combined analysis,
we allowed a galaxy-to-mass bias in the HTP spectrum to vary:
\begin{equation}
  \label{b}
  b_{\rm I(HTP)}=\left[
      \frac{{\cal P}_{\rm I(HTP)}(k)}{{\cal P}_{\rm (HTP)}(k)}
    \right]^{1/2},
\end{equation}
while fixing the amplitude of ${\cal P}_{\rm (PD)}(k)$.
The best fit of HTP+PD showed (at 95 per cent confidence level)
\begin{equation}
  \label{b_IHTP}
  b_{\rm I(HTP)}=1.28^{+0.21}_{-0.18}.
\end{equation}
This value also implies the bias between the HTP IRAS galaxy power spectrum
and the PD mass power spectrum.
We note in figure~\ref{figure1} that
for a direct comparison,
we have rescaled down
the amplitudes of the HTP data (dots with crosses)
and their best fit (the dashed line)
by a factor of $b_{\rm I(HTP)}^2=1.28^2$.

The estimates of $\Gamma$ and $n$ from the HTP+PS are shown in table \ref{table1}.
The corresponding best fit of ${\cal P}(k)$ 
is plotted as the solid line in figure~\ref{figure1}.
We see that
it is much closer to the best fit of PD (the dotted line),
and this is mainly due to the smaller error bars in the PD.
From the last column of table \ref{table1},
we also see that 
the best fits of each case (HTP, PS, and HTP+PS)
are good, and agree well with each other.
Based on the form (\ref{Gamma}) of  $\Gamma$,
it is worth emphasizing that
our estimates of $\Gamma$ and $n$
are consistent with the recent CMB constraints,
which give
$0.3\gsim \Gamma\gsim 0.1$ and $n\approx 1$
\cite{ma1,b98,maxiboom}.

\begin{figure}
  \centering 
  \leavevmode\epsfxsize=8cm \epsfbox{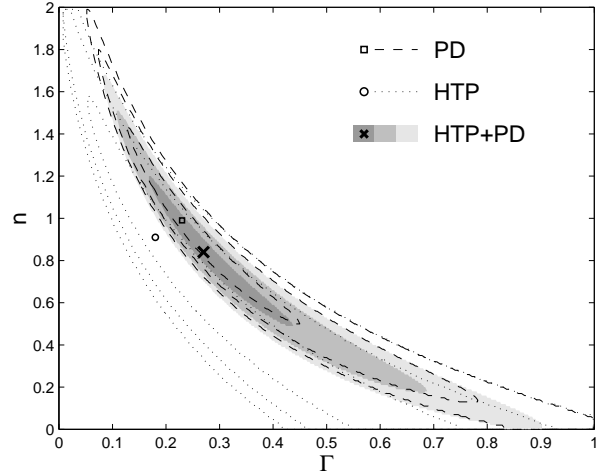}\\
  \caption[]
  {The 68\%, 95\%, and 99\% contours (inner out)
   of the maximum-likelihood analysis for the parameters $\Gamma$ and $n$,
   based on the PD (dashed), HTP (dotted), and HTP+PD (shaded).
   The strong degeneracy between $\Gamma$ and $n$ is apparent.
   The square, circle, and cross label the best fit of
   each case shown in table~\ref{table1}.}
  \label{figure_Gamma_n}
\end{figure}

One interesting observation 
in all the above likelihood analyses is that
there is a strong degeneracy between the estimates of $n$ and $\Gamma$:
the larger the $\Gamma$, the smaller the $n$.
Figure~\ref{figure_Gamma_n} shows this degeneracy
in the likelihood plots of the three cases (PD, HTP, and HTP+PD).
For any $n$ within $2\geq n\geq 0$,
we found that
the $\Gamma$ of local maximum likelihood can be well fitted by
\begin{equation}
  \label{Gamma_n}
  \Gamma=
  \cases{
    0.657 \exp(-1.38n-0.02n^6) & for HTP, \cr
    0.912 \exp(-1.38n-0.002n^6) & for PD, \cr
    0.868 \exp(-1.38n-0.002n^6) & for HTP+PD, \cr
  }
\end{equation}
all within 3\% error.
This degeneracy is caused by the fact that
${\cal P}(k)$ is lack of feature except a broad peak,
so that 
in producing the same shape of ${\cal P}(k)$
a larger $\Gamma$ (which tends to shift the peak to the right)
can compensate for a smaller $n$ (which tends to shift the peak to the left).
This is theoretically expected 
as can be seen from equations (\ref{Pk}) and (\ref{Tk}).
Thus such a degeneracy allows us to write
\begin{equation}
  \label{Sk'}
  {\cal P}(k;n,\Gamma) \propto {\cal P}(k;1,\Gamma'),
\end{equation}
with the `degenerated shape parameter'
\begin{equation}
  \label{Gamma'}
  \Gamma'\equiv \Gamma'(n,\Gamma) 
  = \Gamma \exp(1.4n-1.4)
  \approx 0.247 \Gamma \exp(1.4n),
\end{equation}
which equals the shape parameter $\Gamma$ when $n=1$.
We verified that
this parameterization is accurate within 10\% error
(when ${\cal P}(k)$ is normalized at $k=0.2 h$Mpc$^{-1}$)
for $1.4>n>0.5$, $0.6>\Gamma>0.09$, $0.28>\Gamma'>0.15$,
and
$0.01 > k/ (h$Mpc$^{-1})> 0.3$
(the range of linear scales probed by observations).
Using this parameterization,
we obtained the best fits of $\Gamma'$ shown in table \ref{table1}.
We note that
with the degeneracy (\ref{Gamma'})
the best fits of $\Gamma'$ are consistent with those of the $n$ and $\Gamma$.
When comparing the best fits of $\Gamma'$ in HTP, PD, and HTP+PD,
we also note that
the larger values of $\Gamma'$ in the HTP+PD and PD analyses
reflect the previously observed fact that
the peak in ${\cal P}(k)$ is confined to smaller scales (larger $k$)
in these two cases (see figure~\ref{figure1}).
In the rest of our analysis,
we shall use the parameterization (\ref{Sk'}),
and adopt the result of HTP+PD.
Its errors can be well approximated by a log-normal distribution
(at 95 per cent confidence level; c.f., table~\ref{table1})
\begin{equation}
  \label{Gamma_0}
  \Gamma'
  = 0.220^{+0.036}_{-0.031}
  \approx 0.220\times 10^{\pm 0.066}.
\end{equation}

In principle, 
we can also use observations to constrain $\sigma_8$ 
via equation (\ref{sigma_R2}).
However,
this is forbidden by
the large uncertainty in the normalization of mass power spectrum,
mainly resulted from the large uncertainty in the 
understanding of the galaxy to mass bias.
Hence, instead,
we shall use the abundance of X-ray clusters 
to constrain $\sigma_8$,
as is one of the main goals of this paper.
Nevertheless,
here we can still use ${\cal P}_{{\rm I(HTP)}}$ alone
to constrain the $\sigma_8$ for the IRAS galaxies,
namely $\sigma_{8{\rm (I)}}$.
A maximum-likelihood method with equation (\ref{Gamma_0}) as a prior
gives an estimate
\begin{equation}
  \label{sigma_8I}
  \sigma_{8{\rm (I)}}=0.78 \pm 0.03,
\end{equation}
at 68 per cent confidence level,
and $\sigma_{8{\rm (I)}}=0.78 \pm 0.06$ at 95 per cent confidence level.
This result will be used in Section~\ref{discussion-beta}
for the discussion of redshift distortion parameter.

\begin{figure}
  \centering 
  \leavevmode\epsfxsize=8cm \epsfbox{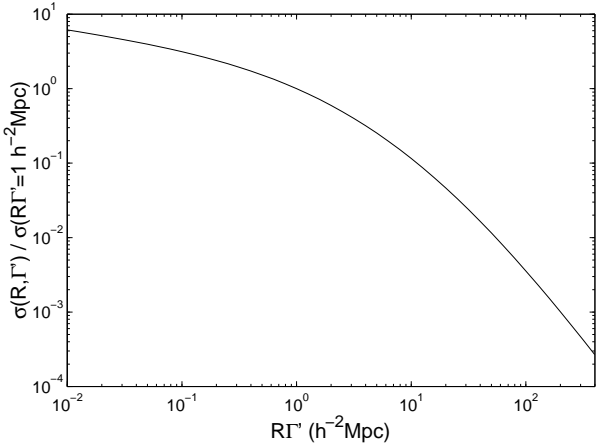}\\
  \caption[]
  {Dependence of amplitude of mass perturbations $\sigma$ on
   the smoothing scale $R$ and the degenerated shape parameter $\Gamma'$.}
  \label{figure2}
\end{figure}

We now investigate the scale dependence of the amplitude of matter perturbations.
By using equation (\ref{sigma_R2}) 
together with equations (\ref{Pk}) and (\ref{Tk}),
we obtained a numerical fit
\begin{equation}
  \label{sigma_R0}
  \sigma (R; n=1,\Gamma)=
  \sigma_{8}
  \frac{\varrho(R\Gamma)}{\varrho(8\Gamma)},
\end{equation}
where $\sigma_{8}\equiv \sigma(R=8)$, and
\begin{equation}
  \label{varrho} 
  \varrho(r) = r^{-\psi(r)},\,~~
  \psi(r)=0.3+\frac{1.45}{1+\left(20/r\right)^{0.35}}.
\end{equation}
Alternatively, we have
\begin{equation}
  \label{R_sigma}
  R (\sigma; n=1,\Gamma)=
  \Gamma^{-1}{\zeta(\frac{\sigma}{\sigma_8}\varrho(8\Gamma))},
\end{equation}
where
\begin{equation}
  \label{zeta} 
  \zeta(s) \equiv \varrho^{-1} \approx 5.2 s^{-0.53} \exp\left(\frac{-s^{0.63}}{0.6}\right).
\end{equation}
Here $\varrho^{-1}$ means the inverse function of $\varrho$,
which is defined in equation (\ref{varrho}).
These fits (eqs.~[\ref{sigma_R0}] and [\ref{R_sigma}]) are accurate within $5\%$ error for
$1000>R\Gamma>0.01$.
We also note that the coupling of $R$ and $\Gamma$ in these fits
is theoretically required 
(see eqs.~[\ref{sigma_R2}] and [\ref{Tk}] with the fact that $q=k/\Gamma$).
To account for the dependence of $\sigma(R)$ on $n$,
we can simply replace the $\Gamma$ in equation (\ref{sigma_R0})
with $\Gamma'$ (see also eq.~[\ref{Sk'}] and context), i.e.,
\begin{equation}
  \label{sigma_R'}
  \sigma(R;n,\Gamma)=\sigma(R; n=1,\Gamma')=\sigma_8
  \frac{\varrho(R\Gamma')}{\varrho(8\Gamma')}.
\end{equation}
The dependence of $\sigma$ on $R$ and $\Gamma'$ 
is shown in figure~\ref{figure2}.

In the later discussion,
we shall also need the red-shift dependence of $\sigma$
for a given present value in a given background cosmology.
This can be described accurately by
\begin{equation}
  \label{sigma_Rz}
  \sigma (R,z)=\sigma (R,z=0)  
  \frac{g(\Omega_{\rm m},\Omega_{\Lambda})}
  {(1+z)g(\Omega_{\rm m0},\Omega_{\Lambda0})},
\end{equation}
where 
$z$ is the red-shift,
$\Omega_\Lambda$ is the energy density of the cosmological constant $\Lambda$,
and \cite{CarPre}
\begin{equation}
\label{g_O}
  g(\Omega_{\rm m},\Omega_\Lambda)  = 
  \frac{2.5\Omega_{\rm m}}
  {\left[\Omega_{\rm m}^{4/7}-\Omega_{\Lambda } +
      (1+{\Omega_{\rm m} / 2})(1+{\Omega_{\Lambda }/70})\right]},
\end{equation}
which accounts for the discrepancy 
of the growing behavior of matter perturbations from that of a 
critical-density universe.
An analytical expression for the evolution of $\Omega_{\rm m}$ is
\begin{equation}
\label{Omegam_Z}
  \Omega_{\rm m} \equiv \Omega_{\rm m}(z)
   =  \frac{\Omega_{\rm m0}(1+z)^3}{(1+z)^2
    (1+\Omega_{\rm m0} z-\Omega_{\Lambda0})+\Omega_{\Lambda0}}.
\end{equation}
Throughout this paper,
we shall investigate open models
with $\Omega_{\rm m}< 1$ and $\Omega_\Lambda=0$
(denoted as OCDM hereafter),
and flat models 
with $\Omega_{\rm m}+\Omega_\Lambda=\Omega_{\rm m0}+\Omega_{\Lambda0}=1$
(denoted as $\Lambda$CDM hereafter).

\section{The mass function}
\label{the-mass-function}

To apply constraint from the observed cluster abundance
onto the amplitude of matter perturbations $\sigma(R)$,
we first need to relate it to the number density of clusters.
By definition, 
the fraction of the total mass within 
collapsed objects larger than a given mass $M$ at a red-shift $z$ is
\begin{equation}
  {\cal F}(M(R,z),z)
  =
  {{\Omega_{\rm m}(> M(R,z),z)} \over {\Omega_{\rm m}(z)}}.
  \label{F_M}
\end{equation}
Here $M(R,z)$ is the cluster mass of a corresponding scale $R$ 
and is defined by 
\begin{equation}
  M(R,z)=\frac{4}{3} \pi R^3 \overline{\rho}(z),
  \label{M}
\end{equation}
where $\overline{\rho}(z)$ is the average energy density of the universe 
at red-shift $z$.
With the numbers given in \cite{KolbTurner},
we obtain
\begin{equation}
  M(R,0)=1.162\times 10^{12} R^{3} \Omega_{\rm m0} h^{-1}{\rm M}_\odot,
  \label{M_R}
\end{equation}
where $R$ is in $h^{-1}$Mpc
and ${\rm M}_\odot$ is the solar mass.
Thus the differential number density of clusters
at a mass interval $dM$ about $M$
can be derived as:
\begin{equation}
  \label{n_Mz}
n_{i}(M,z) dM = -F_i(\mu) \frac{\overline{\rho}}{M\sigma} \frac{d\sigma}{dM} dM,
\quad i=\textrm{PS, ST, or LS},
\end{equation}
where $\sigma\equiv\sigma(R,z)$,
$\overline{\rho}\equiv\overline{\rho}(z)$,
$\mu\equiv \mu(\sigma)$,
and 
\begin{equation}
  \label{F_i}
  F_i(\mu)=\frac{d{\cal F}(M(R,z),z)}{d{\rm ln}\sigma(R,z)}.
\end{equation}
The $n_{i}$ in equation (\ref{n_Mz}) is normally referred as the `mass function'.
In different models of mass function,
the $F_i(\mu)$ (and therefore $n_{i}$) has different forms.
Here we shall consider the models
by Press \& Schechter (1974; hereafter PS),
by Sheth \& Tormen (1999; hereafter ST),
and
by Lee \& Shandarin (1999; hereafter LS):
\begin{equation}
  F_{\rm PS}(\mu) = \sqrt{\frac{2}{\pi}}\mu\exp\Big{(}-\frac{\mu^2}{2}\Big{)},
  \label{F_PS}
\end{equation}
with $\mu=\delta_{\rm c}/{\sigma}$ and $\delta_{\rm c}=1.68647$;
\begin{equation}
  F_{\rm ST}(\mu) = 
  0.322\sqrt{\frac{2}{\pi}}\left(1+\frac{1}{\nu^{0.6}}\right)\nu
  \exp\Big{(}-\frac{\nu^2}{2}\Big{)},
  \label{F_ST}
\end{equation}
with $\nu=\sqrt{0.707}\mu$, $\mu=\delta_{\rm c}/{\sigma}$,
and $\delta_{\rm c}=1.68647$;
and
\begin{eqnarray}
  F_{\rm LS}(\mu) = \frac{25\sqrt{10}}{2\sqrt{\pi}}\mu
\Bigg{[}
\Big{(}\frac{5\mu^2}{3}-\frac{1}{12}\Big{)}
\exp\Big{(}-\frac{5\mu^2}{2}\Big{)}
\times
\nonumber\\
{\rm erfc}\Big{(}{\sqrt{2}\mu}\Big{)}+
\frac{\sqrt{6}}{8}
\exp\Big{(}-\frac{15\mu^2}{4}\Big{)}
\times
\nonumber \\
{\rm erfc}\Big{(}\frac{\sqrt{3}\mu}{2}\Big{)}
-\frac{5\sqrt{2\pi}\mu}{6\pi}
\exp\Big{(}-\frac{9\mu^2}{2}\Big{)}
\Bigg{]},  
  \label{F_LS}
\end{eqnarray}
with $\mu=\lambda_{\rm 3c}/{\sigma}$ and $\lambda_{\rm 3c} = 0.37$.

The $n_{\rm\scriptscriptstyle PS}$ is based on the Press-Schechter formalism,
which relates the mass fraction of collapsed 
objects whose mass is larger than some given threshold $M$,
with the fraction of space 
in which the evolved linear density field exceeds some threshold 
$\delta_{\rm c}$. 
It has been extensively tested against N-body simulations 
with considerable success,
but recently found to have steeper shape 
tilted towards larger mass $M$.
This is mainly due to the spherical-collapse assumption in the PS formalism
(while simulations showed that non-spherical collapses are important
in the clustering process),
and the fact that peaks in a linear field are poorly correlated with
the final spatial locations of dark halos
(for a review, see Lee \& Shandarin 1999 and references therein).
The $n_{\rm\scriptscriptstyle ST}$ by Sheth \& Tormen (1999)
was first obtained as a numerical fit to simulations.  
Sheth, Mo \& Tormen (1999) then showed that
it could be associated with a model of 
ellipsoidal, rather than spherical collapse.
The $n_{\rm\scriptscriptstyle LS}$ by Lee \& Shandarin (1999)
is a full analytical alternative of $n_{\rm\scriptscriptstyle PS}$.
It incorporates non-spherical dynamics
and
assumes that dark halos form in the local maxima of the smallest eigenvalue of
the deformation tensor $\lambda_{\rm 3c}$.
It has been shown to be in better agreement with simulations 
than the PS formalism \cite{LeeShandarin2}.

\begin{figure}
  \centering 
  \leavevmode\epsfxsize=8cm \epsfbox{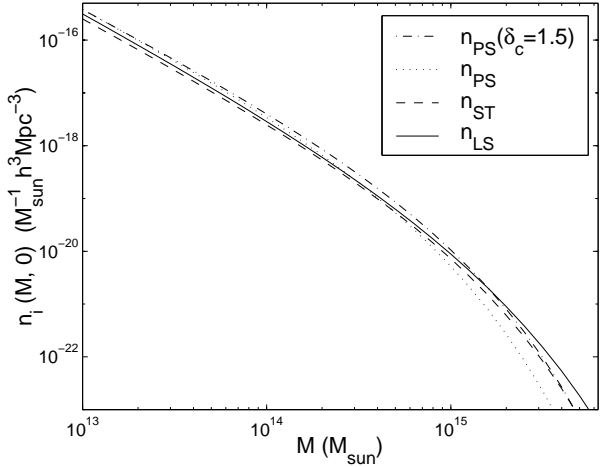}\\
  \caption[]
  {The mass function $n_i$
   (the differential number density
   of collapsed objects within a mass interval $dM$ about $M$)
   of different models ($i=$PS, ST, and LS).
   See equation (\ref{n_Mz}) and context for details.}
  \label{figure3}
\end{figure}

Figure~\ref{figure3} shows the mass functions $n_i$ of these three models.
It indicates that
the PS mass function (dotted line)
predicts fewer massive clusters and more light clusters
than the ST and LS mass functions (dashed and solid lines respectively).
In the literature,
it has been shown that
the PS formalism with a smaller $\delta_c$
can produce more massive clusters and thus resemble the simulations better for large $M$,
but still suffers from the overestimate of light objects
\cite{Tormen,LeeShandarin2}.
To show this effect,
in figure~\ref{figure3} we plotted another PS mass function 
with $\delta_c=1.5$ (dot-dashed line) for comparison.
In addition,
we should also note that
the $\delta_c$ is fixed by the fit to the global mass function,
and therefore not a free parameter essentially.

Finally,
substituting equation (\ref{sigma_R'}) together with (\ref{varrho})
into (\ref{n_Mz})
(see also eq.~[\ref{sigma_Rz}]), 
we obtain the present cluster abundance as
\begin{equation}
  \label{n_i}
  n_i (M, 0) =
  F_i(\mu) \frac{\overline{\rho}}{3M^2}
  \left\{
    \gamma(R\Gamma) +
    \frac{
    1.45 (R\Gamma)^{0.35}\log(R\Gamma)}{[(R\Gamma)^{0.35}+2.85]^2}
  \right\}.
\end{equation}
We note that the $\sigma_8$ dependence of $n_i (M, 0)$ here
has now been implicitly embedded into the $\mu(\sigma)$ of $F_i(\mu)$,
as well as the $M$ and $R$ through equations (\ref{R_sigma}) and (\ref{M}).

\section{Formation redshift of clusters}
\label{formation-redshift}

While 
the cluster abundance in our theoretical formalism (eq.~[\ref{n_i}])
is a function of the cluster mass,
the observed abundance of X-ray clusters is normally in the form
of a function of the X-ray temperature.
Therefore,
we need to relate the X-ray temperature $T$ of a cluster 
with its virial mass $M_{\rm v}$. 
Since clusters of the same mass may have formed 
at different red-shifts resulting in different present temperatures,
we first need to decompose the present number density of clusters
of given mass
into contributions from different formation time,
and then relate the virial mass of a formation time
with the present temperature.
That is,
we first need to find out how much of the present abundance $n_i(M,z=0)$
of given mass $M$ was formed at a given redshift $z$,
and then associate this abundance
with the temperature that corresponds to the given $M$ and $z$.
In this section,
we shall discuss the first step,
while leaving the second step to the next section.

To achieve the first step,
we need to know the probability that
a cluster of given present mass was formed at the given redshift.
Lacey and Cole constructed a merging history for dark matter halos based 
on the excursion set approach and obtained an analytical expression for the 
probability that a galaxy cluster with present virial mass $M$ would have
formed at the given red-shift $z$ (Lacey \& Cole 1993, 1994).
Here we shall generalize their formalism in the following way 
to incorporate different models of mass function.
At first, 
the probability that the formation time of a halo of mass $M_0$ at $z_0$
was earlier than $z$
equals
the probability that it had a parent of mass $M> f M_0$ at redshift $z$,
where $f$ is the fraction of the cluster mass assembled by redshift $z$.
This probability can be easily obtained by a halo counting argument
(modified from Lacey \& Cole 1993):
\begin{equation}
  \label{P0}
  P(M> f M_0, z|M_0,z_0)
    = \int_{S_0}^{S_f} \frac{M_0}{M(S)}
      \frac{F_i(\mu)}{2(S-S_0)} dS,
\end{equation}
where
$S \equiv S(M) = \sigma^2(M,z=0)$, $S_0 \equiv S(M_0)$, $S_f \equiv S(fM_0)$,
and
$\mu=\delta_{\rm c}[\sigma(M,0)/\sigma(M,z)-1] / \sqrt{S-S_0}$
(the $\delta_{\rm c}$ here should be replaced with $\lambda_{\rm 3c}$
for the LS formalism,
and the same applies to the later appearance of $\delta_{\rm c}$
in this section).
Therefore
the probability that a galaxy 
cluster with present virial mass $M$ was formed at a given 
$z$ can be obtained as:
\begin{equation}
  \label{pz0}
  p_z(z)=-\frac{\partial P(M> f M_0, z|M_0,z_0)}{\partial z}.
\end{equation}
With the change of variables
\begin{equation}
  \label{s_w}
  s=\frac{S-S_0}{S_f-S_0}, \quad
  {\omega}
  = \frac{\delta_{\rm c}[\sigma(M,0)/\sigma(M,z)-1]}{\sqrt{S_f-S_0}},
\end{equation}
equation (\ref{pz0}) can be rewritten as
\begin{equation}
  \label{pz}
  p_z(z)=p_\omega(\omega)\frac{\partial \omega}{\partial z},
\end{equation}
where
\begin{equation}
  \label{pw}
  p_\omega(\omega)= -\frac{\partial P}{\partial \omega}
    = - \int_0^1 \left[\frac{R_0}{R(\sigma)}\right]^3
        \frac{\partial F_i(\mu)}{\partial \mu} \frac{ds}{2s^{3/2}},
\end{equation}
and 
\begin{equation}
  \label{mu}
  \mu = \frac{\omega}{\sqrt{s}}, \quad
  \sigma\equiv \sigma(M,z=0)=\left[s(S_f-S_0)+S_0\right]^{1/2}.
\end{equation}
Here
$R_0$ can be obtained from equation (\ref{M_R})
by taking $M=M_0$,
and
$R(\sigma)$ is given 
either
by equation (\ref{R_sigma}) for inflationary CDM models,
or by $R\propto \sigma^{-2/(n_{\rm s}+3)}$ for power-law-spectrum models with
$P(k)\propto k^{n_{\rm s}}$.
In the latter case,
we have $\left[R_0/R(\sigma)\right]^3 = 
\left[1+s(f^{-(n_{\rm s}+3)/3}-1)\right]^{3/(n_{\rm s}+3)}$.
Under certain conditions,
equation (\ref{pw}) can be analytically integrated to yield an
explicit expression of $p_\omega(\omega)$.
For example,
for power-law-spectrum models with $n_{\rm s}=0$,
we have simply $\left[R_0/R(\sigma)\right]^3 = 1+s(f^{-1}-1)$.
Thus we obtain for PS
\begin{eqnarray}
  \begin{array}{ll}
    p_\omega(\omega) 
    = & 2 \omega(f^{-1}-1){\rm erfc}\left({\omega \over \sqrt{2}}\right)- \\
    & {\sqrt {2 \over \pi}}(f^{-1}-2)\exp\left(-{{\omega^2} \over 2}\right),
  \end{array}
  \label{p_w_PS}
\end{eqnarray}
and 
the result for ST can be also straightforwardly obtained.

\begin{figure}
  \centering 
  \leavevmode\epsfxsize=8cm \epsfbox{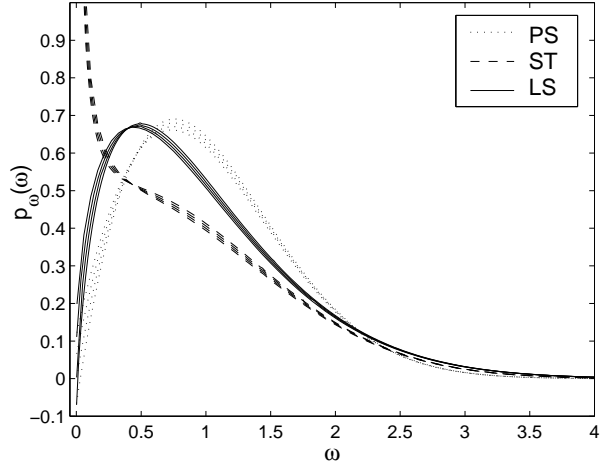}\\
  \caption[]
  {The probability distribution $p_\omega(\omega)$ 
   of halo formation epochs $\omega(z)$
   calculated from equation (\ref{pw})
   of different models,
   for power-law spectra with 
   $n_{\rm s}=-2$, $-1$, $0$ and 1
   (from bottom to top at $\omega=1$).
   For a direct comparison,
   we have rescaled 
   the $p_\omega(\omega)$ of ST with $\lambda_{\rm 3c}/\delta_{\rm c}$,
   and its $\omega$ with $\delta_{\rm c}/\lambda_{\rm 3c}$,
   due to the fact that $\omega\propto \delta_{\rm c}$, $\lambda_{\rm 3c}$.
   Here we have used $f=0.5$.
   We also note that 
   $\omega(z) \propto z$ when $\Omega_{\rm m0}=1$ and $\Lambda=0$.
  }
  \label{figure-pw}
\end{figure}

The form of equation (\ref{pw}) is general
and thus suitable for any given model of mass function.
Once given the mass function,
which is specified by $F_i(\mu)$ (see eq.~[\ref{n_Mz}]),
we can first use equation (\ref{pw}) to obtain $p_\omega$,
and then equation (\ref{pz}) to get $p_z(z)$.
From equations~(\ref{s_w}), (\ref{pz}),  and (\ref{sigma_Rz}),
we first note that for $\Omega_{\rm m0}=1$ and $\Lambda=0$, 
$\omega(z)$ is proportional to $z$,
so that the conversion from $p_\omega(\omega)$ to $p_z(z)$
requires only a linear rescale in both $\omega$ and $p_\omega(\omega)$.
Even for other values of $\Omega_{\rm m0}$ and $\Lambda$,
which result in non-linear conversion from $\omega$ to $z$ 
and thus from $p_\omega(\omega)$ to $p_z(z)$,
the relative ratios between the $p_z(z)$ of the PS, ST and LS formalisms
will remain the same as in the $p_\omega(\omega)$,
i.e. the conversion from $p_\omega(\omega)$ to $p_z(z)$
is independent of the choice of the mass function.
This allows us to directly compare the $p_\omega(\omega)$ of PS, ST and LS,
and keep the conclusions still available for $p_z(z)$.
Figure~\ref{figure-pw} shows the $p_\omega(\omega)$ in 
the PS (dotted lines), ST (dashed lines), and LS (solid lines)
formalisms,
for power-law-spectrum models with $n_{\rm s}=-2,-1,0$ and $1$.
For a direct comparison,
we have rescaled 
the $p_\omega(\omega)$ of ST with $\lambda_{\rm 3c}/\delta_{\rm c}$,
and its $\omega$ with $\delta_{\rm c}/\lambda_{\rm 3c}$,
due to the fact that $\omega\propto \delta_{\rm c}$, $\lambda_{\rm 3c}$
(see eq.~[\ref{s_w}]).
We have also verified that
for the inflationary CDM models,
the $p_\omega(\omega)$ (and thus $p_z(z)$) 
of a chosen mass function
approximately interpolates between those based on
the pow-law-spectrum models.
We should also notice that
the PS results here (the dotted lines) are identical to those
presented in Lacey \& Cole (1993).

In comparing the results of PS, ST and LS in figure~\ref{figure-pw},
we see that
the $p_\omega(\omega)$ of non-spherical-collapse models (ST and LS)
have larger tails at high $\omega$,
allowing clusters to form at earlier time.
This is consistent with the argument that
in non-spherical-collapse models,
halos are considered formed as long as one of three major axes has collapsed,
leading to the earlier formation of clusters.
This earlier formation of clusters consumes more over-density regions
at early time than in the PS formalism,
so as to cause the delay of the active formation epoch,
i.e. the maximum of $p_\omega(\omega)$ being located at a lower $\omega$
(see the solid and dashed lines in figure~\ref{figure-pw}).
We also note that the dramatic increase of $p_\omega(\omega)$ towards
$\omega=0$ in the ST model is nearly divergent,
and this is caused by the mathematical form of $F_{\rm ST}$.

To implement all the analysis properly,
we shall use the numerical result of equation (\ref{pw})
based on the inflationary models (invoking eq.~[\ref{R_sigma}]).
The value of $f$ we shall use is \cite{VL2,Navarro} (at $95$\% confidence interval)
\begin{equation}
  \label{f_0}
  f=0.75\pm 0.15.
\end{equation} 
Thus the present number density of clusters of present mass $M$ 
that were formed at red-shift $z$ can now be obtained as
$n_i(M,0)p_z(z){\rm d}M{\rm d}z$.

\section{Mass-temperature relation and X-ray cluster abundance}
\label{Mass-temperature}

We then need to associate the abundance $n_i(M,0)p_z(z){\rm d}M{\rm d}z$ 
with the temperature 
that corresponds the given $M$ and $z$.
This requires the knowledge about the relation between 
the virial mass $M\equiv M_{\rm v}$ of a formation red-shift $z$
and its present temperature $T$.
Here
we shall use the result of Viana \& Liddle (1996, 1999),
which was calibrated to 
the hydrodynamical $N$-body simulations of White et al.\ (1993b),
and also shown to agree well with the result by Bryan \& Norman (1998).
Their result of the $M$-$T$ relation is (modified from Viana \& Liddle 1999):
\begin{equation}
  \label{M_v}
  \begin{array}{l}
    M_{\rm v}=(1.23\pm0.33)\times 10^{15}
    \left[
      \frac{\Omega_{\rm mt}^{b(\Omega_{\rm mt})}}
      {\Omega_{\rm m}^0}
    \right]^{1/2}
    \left[
      \frac{1.67}{1+z_{\rm t}}\times \right. \\
    ~~~~~~~~~\left.
    {2(2-\eta)(4-\eta)^2
        \over
        64-56\eta+24\eta^2-7\eta^3+\eta^4
        }
      \frac{k_{\rm B}T}{7.5{\rm keV}}
    \right]^{3/2}
    h^{-1}{\rm M}_\odot,
  \end{array}
\end{equation}
where
the error is 1-sigma,
$z_{\rm t}$ is the turnaround red-shift,
$\Omega_{\rm mt}\equiv \Omega_{\rm m}(z_{\rm t})$,
and
\begin{eqnarray}
  \label{eta}
  \eta & \equiv & \eta(z_{\rm t}) = 
  \frac{32}{9\pi^2}
  \frac{\Omega_{\Lambda}^0 \Omega_{\rm mt}^{b(\Omega_{\rm mt})}}
  {\Omega_{\rm m}^0(1+z_{\rm t})^3},
  \\
  b(\Omega) & = & \left\{
      \begin{array}{ll}
        0.76-0.25\Omega & ({\rm OCDM}),
        \\
        0.73-0.23\Omega & ({\rm \Lambda CDM}).
      \end{array}
    \right.
\end{eqnarray}
For a given red-shift $z\equiv z_{\rm c}$ of cluster collapse,
the turnaround red-shift $z_{\rm t}$ is easily obtained using the 
fact that $2t(z_{\rm t})=t(z_{\rm c})$, 
where
\begin{equation}
  t(z)=
  \cases{
    \frac{2}{3}H_0^{-1}(1+z)^{-3/2} 
     \quad \textrm{for $\Omega_{\rm m0}=1$, $\Omega_{\Lambda 0}=0$}, \cr
    \frac{H_0^{-1}\Omega_{\rm m0}}{2(1-\Omega_{\rm m0})^{3/2}}
    \left[
    \frac{2(1-\Omega_{\rm m0})^{1/2}(1+z\Omega_{\rm m0})^{1/2}}
         {\Omega_{\rm m0}(1+z)}\right. \cr
    \left.-\cosh^{-1}\left(\frac{z\Omega_{\rm m0}-\Omega_{\rm m0}+2}{z\Omega_{\rm m0}+\Omega_{\rm m0}}\right)
    \right]
    \quad \textrm{for OCDM},\cr
    \frac{2H_0^{-1}}{3\Omega_{\Lambda 0}^{1/2}}
    \ln\left\{
       \frac{\Omega_{\Lambda 0}^{1/2}+[\Omega_{\Lambda 0}+\Omega_{\rm m0}(1+z)^3]^{1/2}}{\Omega_{\rm m0}^{1/2}(1+z)^{3/2}}
       \right\}
    \; \textrm{for $\Lambda$CDM}.\cr
  }
\end{equation}
We note that
although other work of the $M$-$T$ relation may suggest
a different normalization from the one used in equation (\ref{M_v}),
the error quoted here should have reasonably included the possible deviations.
For example,
the normalization used by Pierpaoli, Scott \& White (2000)
is about 15\% lower than here,
and this is well within our 1-sigma,
which is about 27\% of the central value.
In addition,
we shall also investigate the dependence of our final result on this,
so that one can easily extrapolate our final result
for different $M$-$T$ normalization (see section~\ref{uncertainties}).

Putting all the above results together,
we can now estimate the present  number density of 
galaxy clusters that were formed at a given red-shift $z$ with a mean 
X-ray temperature $k_{\rm B}T$: 
\begin{equation}
  \label{hat_n}
  \hat{n}_i(T,z){\rm d}(k_{\rm B}T){\rm d}z
  =
  n_i(M,0)p_z(z)\frac{3M}{2k_{\rm B}T}{\rm d}(k_{\rm B}T){\rm d}z,
\end{equation}
where we have used the fact that
${\partial M}/{\partial (k_{\rm B}T)}={3M}/{2k_{\rm B}T}$.
By integrating this over 
the temperature $T$ and the formation redshift $z\equiv z_{\rm c}$,
we obtain the theoretically predicted abundance of X-ray clusters
observed at $z_{\rm obs}$
with a temperature greater than a given threshold $T_{\rm th}$:
\begin{equation}
  \label{N_thy}
  N_{\rm thy}(>T_{\rm th}, z_{\rm obs}) =
  \int_{0}^{\infty} \int_{k_{\rm B}T_{\rm th}}^{\infty}
  \hat{n}_i(T,z){\rm d}(k_{\rm B}T){\rm d}z.
\end{equation}
We note that
the $z_{\rm obs}$ on the left hand side means completely differently
from the $z$ on the right hand side.
The former means the red-shift of the observed clusters,
while the latter means the formation red-shift of those clusters relative to $z_{\rm obs}$.
Thus we need to integrate the above equation as if we were placed at $z_{\rm obs}$,
i.e.\ we need to first shift all relevant physical conditions today
($\Omega_{\rm m0}$, $\Omega_{\rm \Lambda 0}$, etc.)
to the epoch $z_{\rm obs}$ 
($\Omega_{\rm m}(z_{\rm obs})$, $\Omega_{\rm \Lambda}(z_{\rm obs})$, etc.),
and then integrate equation (\ref{N_thy}) as if $z_{\rm obs}=0$.

A comparison between the observed 
cluster abundance and the above theoretical prediction will give us an estimate 
of $\sigma_8$. 
The observed cluster abundance we shall use for this comparison is
that given by Viana \& Liddle \shortcite{VL2}, based on the dataset in
Henry \& Arnaud \shortcite{HA}, and updated by Henry \shortcite{Henry1}.
It is an abundance at $z_{\rm obs}=0.05$ 
with X-ray temperature exceeding $6.2  {\rm keV}$:
\begin{equation}
  N_{\rm obs}(>6.2 {\rm keV}, 0.05)
  =1.53\times 10^{-7\pm 0.16}
  h^3{\rm Mpc}^{-3}.
  \label{N_obs}
\end{equation}
The uncertainty in (\ref{N_obs}) is the 1-sigma interval,
and has taken into account the effect of
temperature measurement errors. The reasons for concentrating on 
galaxy clusters with temperature larger than $6.2 {\rm keV}$ has been extensively discussed by Viana \& Liddle \shortcite{VL2}.


\section{Results and discussion}
\label{results-and-discussion}

\subsection{The $\sigma_8$}
\label{the_sigma_8}

\begin{figure}
  \centering 
  \leavevmode\epsfxsize=8cm \epsfbox{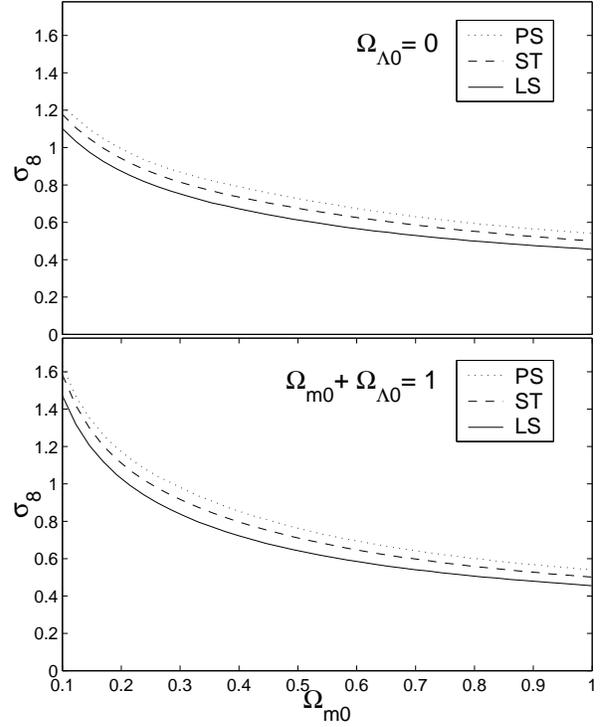}\\
  \caption[]
  {The cluster-abundance-normalized $\sigma_{8}$
    based on different mass functions.
    The upper and lower panels show the results 
    in the OCDM and $\Lambda$CDM models respectively.}
  \label{figure4}
\end{figure}

\begin{figure}
  \centering 
  \leavevmode\epsfxsize=8cm \epsfbox{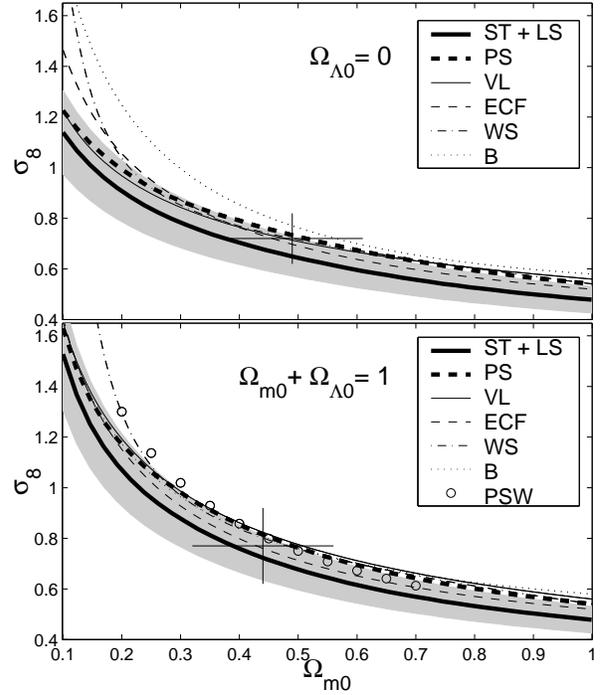}\\
  \caption[]
  {The cluster-abundance-normalized $\sigma_{8}$
    based on ST+LS (thick solid lines) and PS (thick dashed lines),
    compared with results of
    Viana \& Liddle (1999; VL, thin solid),
    Eke, Cole \& Frenk (1996; ECF, thin dashed),
    Wang \& Steinhardt (1998; WS, dot-dashed),
    Borgani et al.\ (1999; B, dotted),
    and
    Pierpaoli, Scott \& White (2000; PSW, circles in the lower panel),
    in the OCDM (upper panel) and $\Lambda$CDM (lower panel) models.
    The shaded areas are the 1-sigma regions of the ST+LS result.
    The crosses are the observational constraints derived
    by Henry (2000).
}
  \label{figure7}
\end{figure}

By comparing the observed cluster abundance (\ref{N_obs})
with the theoretical prediction (\ref{N_thy}),
we obtained the values of $\sigma_8$
based on different models of mass function.
Figure~\ref{figure4} shows the results as functions of
$\Omega_{\rm m0}$ and $\Omega_{\Lambda 0}$.
These results can be fitted by
\begin{equation}
  \label{sigma_8}
  \sigma_{8(i)}(\Omega_{\rm m0},\Omega_{\Lambda 0})
    = c_1 \Omega_{\rm m0}^{\alpha},
    \textrm{  $i=$PS, ST, LS, and ST+LS},
\end{equation}
where
\begin{equation}
  \label{sigma_8_pow}
  \alpha
  \equiv
  \alpha(\Omega_{\rm m0},\Omega_{\Lambda 0}) = 
  -0.3 -0.17 \Omega_{\rm m0}^{c_2} 
  - 0.13 \Omega_{\Lambda 0},
\end{equation}
and the values of $c_1$ and $c_2$
are given in table~\ref{table2}.
These fits are accurate within 2\% error for $1\geq\Omega_{\rm m0}\geq 0.1 $.
We notice that
these results of $\sigma_8$
have only simple dependence on $\Omega_\Lambda$
(the last term of eq.~[\ref{sigma_8_pow}]),
and this dependence is independent on the choice of mass function.
It is clear that for a given $\Omega_{\rm m0}$,
the cosmological constant tends to increase the normalization of $\sigma_8$.
Also given in equation (\ref{sigma_8}) and table \ref{table2}
is the mean of the ST and LS results,
labeled as `ST+LS' (see next section for more details).
It is plotted as the thick solid line in figure~\ref{figure7},
and we shall refer to this as the combined result of non-spherical-collapse models.

\begin{table}
  \centering 
  \begin{tabular}{c|cccc}
    \hline\hline
    $i$ & PS & ST & LS & ST+LS \\
    \hline
    $c_1$ & 0.54 & 0.50  & 0.455 & 0.477 \\
    \hline
    $c_2$  & 0.45 & 0.37  & 0.31 & 0.34  \\
    \hline\hline
  \end{tabular}
  \caption{Values in the fit (\ref{sigma_8}) of $\sigma_8$,
  based on different models of mass function.}
  \label{table2}
\end{table}

Figure~\ref{figure4} indicates that
the $\sigma_{8\rm (ST)}$ and $\sigma_{8\rm (LS)}$ are systematically lower than
$\sigma_{8\rm (PS)}$. 
On average within the ranges of
$\Omega_{\rm m0}$ and $\Omega_{\Lambda 0}$ probed here,
we found that
$\sigma_{8\rm (ST)}$ is 7\% and $\sigma_{8\rm (LS)}$ is 15\% 
smaller than $\sigma_{8\rm (PS)}$.
In figure~\ref{figure7},
we also see that
the combined non-spherical-collapse results $\sigma_{8\rm (ST+LS)}$ (thick solid lines)
are on average 11\% lower than $\sigma_{8\rm (PS)}$ (thick dashed lines).
This implies that
the inclusion of non-spherical collapse considerably reduces
the amplitude of mass perturbations 
required for resembling the observed cluster abundance.
This can be understood as the consequence
of the higher abundance of massive clusters in the non-spherical-collapse models.
In figure~\ref{figure3},
we saw that the non-spherical-collapse models (ST and ST) predict more massive clusters.
Adding the fact that
clusters are relatively rare objects that exist only with high mass,
we know that
the amplitude of matter perturbations required to resemble the
observed cluster abundance will be lower in non-spherical-collapse models,
resulting in the lower $\sigma_8$ 
as seen in figures~\ref{figure4} and \ref{figure7}.

In figure~\ref{figure7},
we have also plotted other results in the literature for comparison.
They are results 
by Viana \& Liddle (1999; VL, thin solid lines),
by Eke, Cole \& Frenk (1996; ECF, thin dashed lines),
by Wang \& Steinhardt (1998; WS, dot-dashed lines),
by Borgani et al.\ (1999; B, dotted lines),
and
by Pierpaoli, Scott \& White (2000; PSW, circles in the lower panel).
All these results were based on the PS formalism
except for the last one (PSW),
which was based on the ST mass function.
We see that for $\Omega_{\rm m0}>0.2$,
our $\sigma_{8 \rm (PS)}$ is consistent well within 10\% error
with all the PS-based results except for the Borgani et al.\ (1999)
in open models.
In particular,
our $\sigma_{8 \rm (PS)}$
is consistent with that of VL within 5\% error.
Since our PS framework is parallel to that of VL,
we verified that 
this small deviation is partly caused by the fact that
we used the inflationary CDM model to calculate $p_z(z)$
(the probability that a cluster of given mass was formed at $z$),
while VL used the $p_z(z)$ based on
a power-law-spectrum model with $n_{\rm s}=0$ (i.e., our eq.~[\ref{p_w_PS}]).
On the other hand,
for non-spherical-collapse models,
our $\sigma_{8 \rm (ST+LS)}$ is lower than that of PSW.
This discrepancy is mainly caused by 
the fact that
we integrated the abundance over the formation shift
while the PSW did not.
We verified that
the ignorance of the integration over the formation shift
will result in a significant overestimate of $\sigma_{8}$,
especially for low $\Omega_{\rm m0}$ cosmologies.
This is because
clusters of the same mass but formed in the past 
will have higher temperature and thus more observable  than those formed today,
so that the required $\sigma_8$ to resemble the observation
in models allowing clusters to form in the past will be lower.
We also note that
our normalization for the $M$-$T$ relation is higher than that of the PSW.
If we adopt their normalization,
our $\sigma_{8 \rm (ST+LS)}$ will be even lowered by about 5\%.
Overall,
it is clear that
our $\sigma_{8 \rm (ST+LS)}$ is lower than any previous results,
and this is mainly attributed to the inclusion of
non-spherical collapse in our formalism.

It should be also noticed that
in Henry (2000),
the maximum likelihood fits of cluster evolution models to 
the observed cluster X-ray temperatures
gave
$\Omega_{\rm m0}=0.49\pm 0.12$ and $\sigma_8=0.72\pm 0.1$ for OCDM,
and 
$\Omega_{\rm m0}=0.44\pm 0.12$ and $\sigma_8=0.77\pm 0.15$ for $\Lambda$CDM,
all at 68 per cent confidence level.
These results are plotted as the crosses in figure~\ref{figure7}.
They are consistent with all the results presented here.

\subsection{Uncertainties}
\label{uncertainties}

To know how stringent our results of $\sigma_8$ are,
we investigate the dependence of $\sigma_8$ 
on $M_{\rm v}$, $f$, $N\equiv N_{\rm obs}(>6.2 {\rm keV}, 0.05)$ and $\Gamma'$,
because they carry uncertainties in our analysis pipeline.
Empirically we found
\begin{eqnarray}
  \frac{\sigma_8(\Gamma',f,M_{\rm v},N)}{\sigma_8(\Gamma'_{(0)},f_{(0)},M_{\rm v(0)},N_{(0)})}
  =
  \left(\frac{\Gamma'}{\Gamma'_{(0)}}\right)^{p_\Gamma}
  \left(\frac{f}{f_{(0)}}\right)^{p_f} \times
  \nonumber \\
  \left(\frac{M_{\rm v}}{M_{\rm v(0)}}\right)^{p_M}
  \left(\frac{N+\tilde{N}}{N_{(0)}+\tilde{N}}\right)^{p_N},
  \label{sigma_8_dep}
\end{eqnarray}
where the subscript $(0)$ denotes the central values 
(see eqs.~[\ref{Gamma_0}], [\ref{f_0}], [\ref{M_v}], and [\ref{N_obs}])
that were used to obtain the main result  (\ref{sigma_8}),
and
\begin{eqnarray}
  p_{\Gamma'} & = & 0.14-0.13\Omega_{\rm m0}^{0.9}+0.1\Omega_{\Lambda 0}^2,
  \label{p_Gamma}\\
  p_f & = & 1.95-1.9 \Omega_{\rm m0}^{0.12}-0.25\Omega_{\Lambda 0},
  \label{p_f}\\
  p_M & = & 0.41-0.06\Omega_{\rm m0}+0.08\Omega_{\Lambda 0},
  \label{p_M}\\
  p_N & = & 0.12\Omega_{\rm m0}^{-0.15}+0.05\Omega_{\Lambda 0}^2
      -\delta_{\rm PS} 0.021\Omega_{\rm m0}^{-0.24},
  \label{p_N}\\
  \tilde{N} & = & 5\times 10^{-8}h^3{\rm Mpc}^{-3},
  \label{hatN}
\end{eqnarray}
with $\delta_{\rm PS}$ equal unity for the PS, and zero otherwise.
Equation (\ref{sigma_8_dep}) is accurate within $3\%$, $3\%$, $2\%$, and $0.5\%$ errors
for 
\begin{eqnarray}
  1.8   > &  \frac{M_{\rm v}}{M_{\rm v(0)}}  & >   0.2, \label{cond_M}\\
  0.9   > &  f  & >   0.53, \label{cond_f}\\
  3   > &  \frac{N}{N_{(0)}}  & >   0.3, \label{cond_N}\\
  0.28   > &  \Gamma'  & >   0.17 \label{cond_Gamma}.
\end{eqnarray}
From equation (\ref{sigma_8_dep}),
we first note that
within the parameter range probed here 
($1\geq \Omega_{\rm m0}\geq 0.1$
and $\Omega_{\Lambda 0}=1-\Omega_{\rm m0}$ or $0$),
the $\sigma_8$ is always a monotonically increasing function
of all $M_{\rm v}$, $N$, $f$ and $\Gamma'$.
Second,
it can be also inferred that
within the $68\%$ confidence level of 
$M_{\rm v}$, $N$, $f$ and $\Gamma'$
(see eqs.~[\ref{M_v}], [\ref{N_obs}], [\ref{f_0}], and [\ref{Gamma_0}]),
the resulting $\sigma_8$ 
within the parameter range probed here 
can vary up to
$14\%$, $6\%$, $5\%$, and $2\%$ respectively,
depending on the background cosmology.
Thus we see that
an improvement in reducing the uncertainty in $M_{\rm v}$,
the normalization of the $M$-$T$ relation,
can most efficiently reduce the uncertainty in the resulting $\sigma_8$.
To the contrary,
a more accurate estimation of the degenerated shape parameter $\Gamma'$
will not change the resulting $\sigma_8$ much.
This observation helps relax the worry about the accuracy in using
the parameterization (\ref{Sk'}), and
the approximation (\ref{Gamma}),
which we have used to model baryonic effects.

To investigate the overall uncertainties in $\sigma_8$
(that is, to include the uncertainties from all
$M_{\rm v}$, $f$, $N$ and $\Gamma'$ simultaneously),
we implemented Monte Carlo simulations,
with the distributions of the uncertainties in
$\Gamma'$ and $N_{\rm obs}(>6.2 {\rm keV}, 0.05)$ as log-normal,
and those in $M_{\rm v}$ and $f$ as Gaussian
(see eqs.~[\ref{Gamma_0}], [\ref{f_0}], [\ref{M_v}], and [\ref{N_obs}]).
We found that
the distribution of the resulting $\sigma_8$ is very much Gaussian,
with a mean given by equation (\ref{sigma_8}) and table \ref{table2},
and a standard deviation ($68\%$ confidence level)
\begin{equation}
  \label{sigma_8_1sigma}
  \epsilon_i(\Omega_{\rm m0},\Omega_{\Lambda 0}) = 
  0.1 \Omega_{\rm m0}^{-0.15}\sigma_{8(i)}(\Omega_{\rm m0},\Omega_{\Lambda 0}),
  \quad i=\textrm{PS, ST, LS}.
\end{equation}
We have verified that
this result is very weekly dependent on $\Omega_{\Lambda 0}$
and the choice of the mass function (PS, ST, or LS),
and these two aspects together contribute only 
a deviation of less than 3\%  from the above result.

To obtain a single result for the non-spherical-collapse models (ST and LS),
we linearly combine the probability distribution functions of 
$\sigma_{8\rm (ST)}$ and $\sigma_{8\rm (LS)}$ 
obtained from the above Monte Carlo simulations.
The resulting distribution of  $\sigma_8$  is very Gaussian.
It has a mean given by equation (\ref{sigma_8}) and table \ref{table2} 
(the ST+LS result),
and a standard deviation
\begin{equation}
  \label{sigma_8_1sigma_STLS}
  \epsilon_{\rm ST+LS}(\Omega_{\rm m0},\Omega_{\Lambda 0}) = 
  0.11 \Omega_{\rm m0}^{-0.12}\sigma_{8{\rm (ST+LS)}}(\Omega_{\rm m0},\Omega_{\Lambda 0}).
\end{equation}
Figure~\ref{figure7} shows this ST+LS result
as the thick solid lines (the mean; eq.~[\ref{sigma_8}] and tab.~\ref{table2}) 
with shaded areas (the 1-sigma regions; eq.~[\ref{sigma_8_1sigma_STLS}]).
Also plotted for comparison is the PS result (the thick dashed lines). 
It is clear that
the result based on non-spherical-collapse models
is about one sigma lower than the PS result,
due to the origin already argued in section~\ref{the_sigma_8}.

In our entire analysis,
we have used $\Omega_{\rm B0}=0.05$,
but verified that for $0.05\geq \Omega_{\rm B0}\geq 0$,
the normalization of $\sigma_8$ was affected by less than 1\%.
We have also tested that
the difference between the $\sigma_8$ 
based on the $p_z(z)$ of the standard CDM model
and of the power-law-spectrum model with $n_{\rm s}=0$
is less than 3\% in all cases.
In addition,
we explore the effect of the integration
over the formation redshift $p_z(z)$ based on
different models of mass function,
i.e.\
while keeping the mass function the same (PS, ST or LS),
we use different $p_{z(i)}(z)$ ($i=$PS, ST or LS) to calculate $\sigma_8$.
It is found that
for a given model of mass function,
the $\sigma_8$ is higher
when using $p_{z \rm (LS)}(z)$
than using $p_{z \rm (PS)}(z)$,
and is even higher when using $p_{z \rm (ST)}(z)$.
Adding the fact that
$p_{z \rm (ST)}(z)$ and $p_{z \rm (LS)}(z)$
have maxima located at lower $z$ than $p_{z \rm (PS)}(z)$
(see figure~\ref{figure-pw}),
the above result is consistent with the previously 
argued and verified fact that
the required $\sigma_8$ to reproduce the observed cluster abundance
will be lower if clusters are actively formed earlier 
(and thus have higher temperature and are more observable).
Nevertheless,
the difference in the $\sigma_8$ using different $p_z(z)$
but the same mass function
is always less than 3\%.
Therefore,
we conclude that
the previously observed lower normalization of $\sigma_8$
in the non-spherical-collapse models is indeed mainly caused by
the form of the mass function itself,
rather than by its form of the  formation-redshift probability,
which in fact has opposite effect.

\subsection{Redshift distortion parameter}
\label{discussion-beta}

\begin{figure} 
  \centering 
  \leavevmode\epsfxsize=8cm \epsfbox{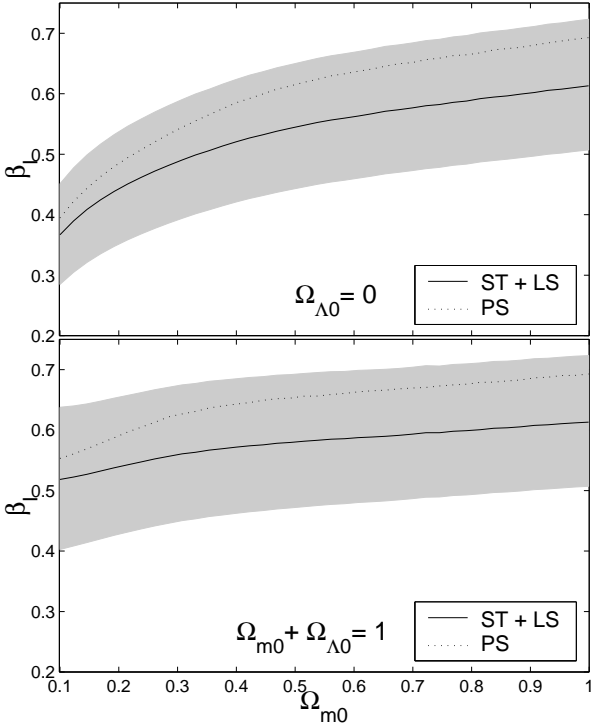}\\
  \caption[]
  {Linear redshift distortion parameter $\beta_{\rm I}$,
   estimated from the combined maximum-likelihood analysis of
   $\sigma_{8{\rm (I)}}$ and $\sigma_8$.
   The solid lines are based on non-spherical-collapse models (ST+LS),
   with the shaded areas representing the regions
   within 68 per cent confidence level.
   The dotted lines are based on the PS formalism.
   }
  \label{figure8}
\end{figure}

Another important aspect in the study of large-scale structure is
the so-called `redshift distortion parameter'.
It quantifies the confusion 
between the Hubble flow and the peculiar velocities,
and is analytically defined as \cite{Kaiser}
\begin{equation}
  \label{beta_I}
  \beta_{j}(\Omega_{\rm m0},\Omega_{\Lambda 0})
	=\frac{f(\Omega_{\rm m0},\Omega_{\Lambda 0})}{b_{j}}
	=f(\Omega_{\rm m0},\Omega_{\Lambda 0}) \frac{\sigma_8}{\sigma_{8(j)}},
\end{equation}
where $b_{j}\equiv \sigma_{8(j)} / \sigma_8$
is the $j$-type galaxy to mass bias ($j=$IRAS, optical, etc.), 
and \cite{Lahav}
\begin{equation}
  \label{f}
  f(\Omega_{\rm m0},\Omega_{\Lambda 0})\approx 
    \Omega_{\rm m0}^{0.6}
    +\frac{\Omega_{\Lambda 0}}{70}(1+\frac{\Omega_{\rm m0}}{2}),
\end{equation}
which is the rate of growth of matter perturbations at the present epoch.
Here we shall estimate $\beta_{\rm I}$,
with the subscript `I' indicating the IRAS galaxies.
To this end,
we can substitute the result (\ref{sigma_8}) (which is calibrated from cluster abundance)
and the result (\ref{sigma_8I}) (which is obtained from the IRAS PSCz survey)
into equation (\ref{beta_I}).
By combining their likelihoods together,
we obtained the maximum-likelihood result,
which can be fitted as
\begin{equation}
  \label{beta_Ifit}
  \beta_{{\rm I} (i)}(\Omega_{\rm m0},\Omega_{\Lambda 0})=
	d_1 \Omega_{\rm m0}^{d_2-0.16(\Omega_{\rm m0}+\Omega_{\Lambda 0})},
  \textrm{$i=$PS, ST+LS,}
\end{equation}
where the values of $d_1$ and $d_2$ are given in table~\ref{table3}.
This fit is accurate within 2\% error.
The errors at 68 per cent confidence level was also found as
\begin{equation}
  \label{beta_1sigma}
  \epsilon_{{\scriptscriptstyle \beta}(i)}(\Omega_{\rm m0},\Omega_{\Lambda 0})
  = 
  0.176 \Omega_{\rm m0}^{-0.11}\beta_{{\rm I} (i)}(\Omega_{\rm m0},\Omega_{\Lambda 0}).
\end{equation}
This fit is accurate within 1\% error.
These results are plotted in figure~\ref{figure8}.

\begin{table}
  \centering 
  \begin{tabular}{c|cccc}
    \hline\hline
    $i$ & PS & ST+LS \\
    \hline
     $d_1$ & 0.693 & 0.613  \\
    \hline
    $d_2$ & 0.26 & 0.24 \\
    \hline\hline
  \end{tabular}
  \caption{Values in the fit (\ref{beta_Ifit}) of $\beta_{{\rm I} (i)}$,
  based on different models of mass function.}
  \label{table3}
\end{table}

From figure~\ref{figure8},
we first note that
the $\beta_{\rm I(ST+LS)}$ is systematically lower than  $\beta_{\rm I(PS)}$.
This is a direct consequence of the fact that $\beta_{\rm I}\propto \sigma_8$,
which is lower in non-spherical-collapse models as previously observed.
We also see that
the inclusion of the cosmological constant tends to increase $\beta_{\rm I}$.
In the past,
$\beta_{\rm I}$ has been measured by many authors using different observations
(for a review, see Strauss \& Willick 1995).
The combined result of these observations has a central value of about
$\beta_{\rm I}\approx 0.7$,
but with a large 68 per cent confidence region of about $\pm 0.4$.
In recent years,
observations seem to favor a lower value.
For example,
$\beta_{\rm I}=0.41^{+0.13}_{-0.12}$ 
by Hamilton, Tegmark \& Padmanabhan (2000),
and 
$\beta_{\rm I}=0.39 \pm 0.12$
by Taylor et al.\ (2000).
If these measures are correct,
then
the PS-based $\beta_{\rm I(PS)}$ (dotted lines in fig.~\ref{figure8})
will be in trouble except in the very low $\Omega_{\rm m0}$ regime,
and 
the non-spherical-collapse models will serve to relax this situation
(solid lines in fig.~\ref{figure8}).


\section{Conclusion}
\label{conclusion}

In this paper,
we first used the observed linear mass power spectra to estimate
the spectral index $n$, the shape parameter $\Gamma$,
the degenerated shape parameter $\Gamma'$,
and the IRAS $\sigma_{8{\rm (I)}}$
(see table~\ref{table1}, eqs.~[\ref{Gamma'}], [\ref{sigma_8I}]).
We then derived
the probability distribution function of cluster formation redshift
for different models of mass function.
We found that 
clusters of the same mass can form earlier
but have a later epoch of active formation 
in the non-spherical-collapse models (ST and LS)
than in the PS formalism.
This is consistent with the observation from numerical simulations.
Based on different models of mass function and 
their associated probability distributions of formation redshift,
we then used the observed cluster abundance to estimate
the amplitude of matter density perturbations 
on the scale of $8 h^{-1}$Mpc, $\sigma_8$,
and the redshift distortion parameter for IRAS galaxies, $\beta_{\rm I}$,
in both OCDM and $\Lambda$CDM cosmologies.
The $\sigma_8$ and $\beta_{\rm I}$ resulted from
non-spherical-collapse models
are systematically lower than those based on the PS formalism.
We showed that
this is mainly owing to the larger mass function
at the high mass end
in the non-spherical-collapse models.
The origins of the uncertainties in our final results
were also investigated separately,
and we found that
the main contribution is from
the uncertainty in 
the normalization of the virial mass-temperature relation.
Therefore we expect that
further improvement in the study of this normalization
will provide us with more stringent constraint 
on both $\sigma_8$ and $\beta_{\rm I}$.
In addition,
since 
we saw significant corrections in the resulting
$\sigma_8$ and $\beta_{\rm I}$
when switching from the conventional Press-Schechter formalism
to the more recent non-spherical-collapse models,
we urge the use of these models in all relevant studies,
especially when we are entering 
the regime of precision cosmology.


\section*{Acknowledgments}

We thank 
Domingos Barbosa,
Marc Davis, Andrew Jaffe, Andrew Liddle, Joe Silk,
Radek Stompor, Naoshi Sugiyama, and Robert Thacker
for useful discussions and comments,
Ravi Sheth for clarifying the use of the ST formalism (\ref{F_ST}),
and Andrew Hamilton and Max Tegmark for providing us with their
PCSz power spectrum.
We acknowledge support from 
NSF KDI Grant (9872979) and
NASA LTSA Grant (NAG5-6552).



\bsp

\label{lastpage}

\end{document}